# Capacitance–Voltage (C-V) Characterization of Graphene-Silicon Heterojunction Photodiodes

Sarah Riazimehr[a], Melkamu Belete[a], Satender Kataria[a], Olof Engström[b],

Max Christian Lemme[a,b*]

[a]RWTH Aachen University, Chair of Electronic Devices, Faculty of Electrical Engineering and Information Technology, Otto-Blumenthal-Str. 2, 52074 Aachen, Germany

[b]AMO GmbH, Advanced Microelectronic Center Aachen (AMICA), Otto-Blumenthal-Str. 25, 52074 Aachen, Germany

*email: max.lemme@eld.rwth-aachen.de

## Abstract

Heterostructures of two-dimensional (2D) and three-dimensional (3D) materials form efficient devices for utilizing the properties of both classes of materials. Graphene/silicon (G/Si) Schottky diodes have been studied extensively with respect to their optoelectronic properties. Here, we introduce a method to analyze measured capacitance-voltage (C-V) data of G/Si Schottky diodes connected in parallel with G / silicon dioxide / Si (GIS) capacitors. We also demonstrate the accurate extraction of the built-in potential ($\Phi_{bi}$) and the Schottky barrier height ($SBH$) from the measurement data independent of the Richardson constant.

**Introduction:**

Heterostructures of two- and three-dimensional materials have been proposed for various technological applications thanks to their simple device architecture, excellent reproducibility and easy fabrication. Heterostructures of Graphene (G) and silicon (Si) can form Schottky diodes and have been demonstrated in different electronic devices, such as chemical and biological sensors[1–3], rectifiers[4], solar cells[5–8] and photodetectors[9–24]. Their thorough characterization is an important subject for the graphene device research community. In particular, it is essential to define precise methods for electrical characterization of the Schottky junctions in order to understand the current transport mechanism in G/Si Schottky diodes, including their similarities and differences with conventional semiconductor technology. To this end, a number of theoretical and experimental studies have been carried out on G/Si Schottky junctions [19–30]. Important diode parameters such as ideality factor ($n$), Schottky barrier height (*SBH)* and series resistance ($R_S$) have been obtained from current-voltage (I-V) and capacitance-voltage (C-V) measurement techniques. Many of these studies have reported varying values for $n$, *SBH* and $R_S$[19–27]. The variations are often attributed to different measurement and fabrication techniques. We have demonstrated in our previous work[22] that in G/Si diodes, the Schottky junction is typically connected in parallel to a graphene / silicon dioxide / silicon area, that have been termed graphene / insulator / semiconductor (GIS) region. This GIS region is essentially a metal-insulator-semiconductor (MIS) capacitor[22]. The detailed role of this parallel GIS capacitor has frequently been neglected when analyzing the electrical characteristics of the G/Si diodes.

In this work, we introduce a method for accurate C-V characterization of G/Si heterojunctions with a parallel GIS capacitor. To evaluate the effect of the GIS capacitor, we performed C-V measurements at small AC signal on G/n-Si Schottky diodes with varying dimensions of the GIS region and on a reference capacitor without Schottky junctions. From these measurements, we propose a methodology to extract the built-in potential ($\mathnormal{\Phi_{bi}}$) and the *SBH*.

## Results and discussions

Fig. 1a shows schematics of two G/n-Si diodes, $D_1$ and $D_2$, with varying GIS areas. Current density-voltage (J-V) characteristics of both diodes under dark condition are displayed in Fig. 1b in a semi-logarithmic scale. In order to calculate the current density, the measured current is normalized to the G/Si Schottky area, which is marked with a dashed red line in Fig. 1c. Both diodes exhibit rectifying characteristics in the dark with rectification ratios up to $3.8 \times 10^4$ and $2.94 \times 10^4$ for $D_1$ and $D_2$, respectively. The basic parameters of the diodes such as *n*, *SBH* and $R_s$ are obtained from the forward J-V characteristics in dark condition using the Cheung method[31]. For $D_1$ and $D_2$, n of 2.08 and 2.3, SBH of 0.76 and 0.79 eV and $R_S$ of 24 and 14 k$\Omega$ have been extracted, respectively. Details of the calculation of the diode parameters are described in our previous work [27] (see Fig. S2 in supporting information therein). It should be noted that the extracted values of *SBH* slightly differ from the true *SBH* of the devices. This is because the theoretical value of the Richardson constant ($A^{**}$) for n-Si (112 Acm$^{-2}$K$^{-2}$) was considered in order to obtain the *SBH*s.[27] For standard Schottky junctions, where a three-dimensional (3D) metal is in contact with Si, the Richardson constant is typically estimated by considering the velocity vector component of electrons normal to the barrier plane and with a kinetic energy greater than the barrier height. For 2D-graphene on silicon, the electron is locked-up between a potential wall, on one side constituted by electron affinity of the graphene layer and, on the other side, by the silicon energy barrier. Trushin suggested that the current across the G/Si barrier can be estimated by considering the tails of electron wave functions pointing into the n-type silicon conduction band.[32] This motivates a much lower effective Richardson constant than that of 3D-metal/Si junctions. It is then clear from the expression for the reverse saturation current of the thermionic emission model[33], that a lower Richardson constant results in a slightly lower value of *SBH*. Thus, in order to obtain a value for *SHB* with high precision, the Richardson constant must be appointed with care. We have therefore performed C-V measurements to obtain *SBH* values of devices independent of the Richardson constant.

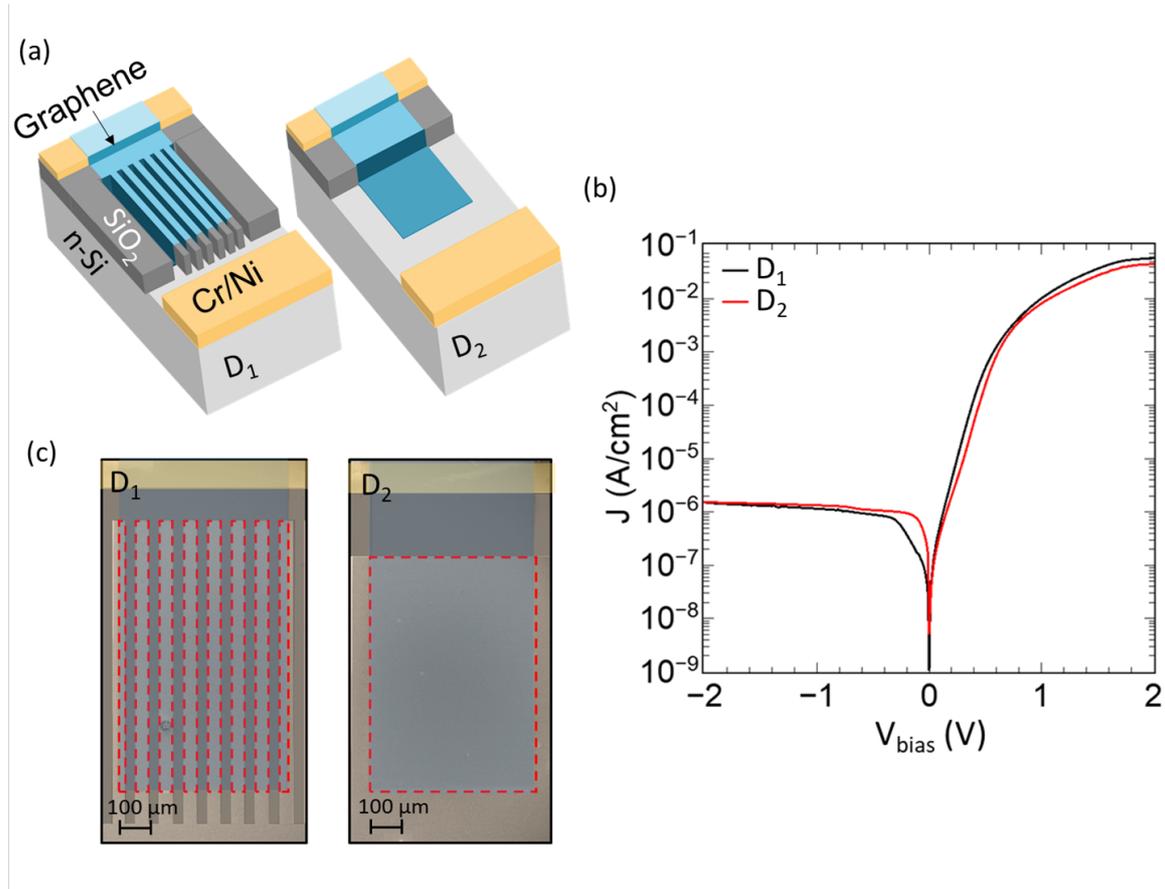

Fig. 1: (a) Schematic of two different G/n-Si diodes, $D_1$ and $D_2$, including a Cr/Ni bulk contact. (b) J-V characteristics of $D_1$ (black line) and $D_2$ (red line) in the dark condition. (c) Scanning electron micrograph (SEM) of the devices with Schottky area marked by red dashed lines. Micrographs have been color enhanced for clarity.

C-V measurements were performed on both devices under study ($D_1$ and $D_2$) and on a test structure ($D_{test}$ – see Fig. 2a) at 10 kHz AC-signal frequency. $D_{test}$ consists of a MIS and a GIS capacitor in a parallel (Fig. 2b). The structure is made of chromium/nickel (Cr/Ni) pad in contact with the graphene electrode, 20 nm thick $SiO_2$ as insulator and n-Si as semiconductor. As shown in Fig. 2a, the MIS bulk contact in $D_{test}$ is located on the top side of the Si chip, similar to $D_1$ and $D_2$. Fig. 2c displays the measured C-V characteristics of $D_{test}$ in comparison with a theoretical one for an ideal MIS capacitor. For an ideal MIS capacitor, the oxide/semiconductor interface charges ($Q_{it}$), the oxide charges ($Q_{ox}$), and the energy difference between the work function of the metal and n-type semiconductor ($Q_{ms}$) are all assumed to be zero. The theoretical C-V plot is calculated using $N_D = 2 \times 10^{15}$ cm$^{-3}$, $t_{ox} = 20$ nm, $\varepsilon_{ox} = 3.9$ and $\varepsilon_S = 11.9$, where $N_D$ is the doping concentration of the Si, $t_{ox}$ is the thickness of the SiO$_2$, $\varepsilon_{ox}$ is the permittivity of the SiO$_2$ and $\varepsilon_S$ is the permittivity of the Si. The curves have been normalized for easier comparison in the voltage range from -4 V to 4 V. This allows extracting the flat-band voltage ($V_{FB}$) and threshold voltage ($V_{th}$ - the voltage at which the inversion happens) for our non-ideal MIS+GIS structure. The procedure further provides information on the presence of imperfections such as trapped charges in the SiO$_2$ and/or at the Si/SiO$_2$ interface. As indicated in Fig. 2c (dashed line), $V_{FB}$ and $V_{th}$ for an ideal MIS structure are 0 V and -0.6 V, respectively. We observed a 0.4 V positive shift for $V_{FB}$ and $V_{th}$ of the test structure, which indicates the presence of trapped negative charges in the SiO$_2$. The observed decrease in the capacitance values of the test structure when the semiconductor is in accumulation originates from the high series resistance due to the MIS bulk contact positioned on the top side of the Si chip and at a fairly long distance from the Si/SiO$_2$ interface.

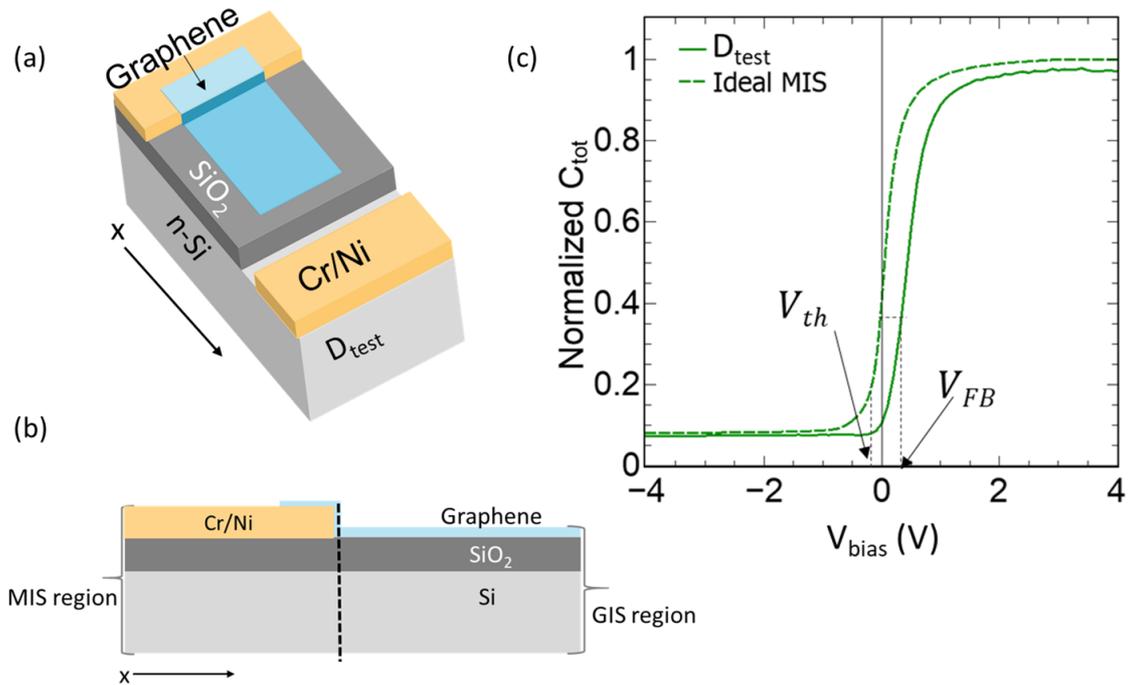

Fig. 2: (a) Schematic of the test capacitor ($D_{test}$). (b) Cross section of the test device showing both MIS and GIS regions. (c) Small-signal C-V characteristics of $D_{test}$ (solid line) compared to a theoretically calculated C-V curve (dashed line) at a frequency of 10 kHz.

Small-signal C−V characteristics of the devices under study (D1 and D2) in comparison with the test device are shown in Fig. 3a. Both $D_1$ and $D_2$ display similar C-V characteristics and can be modelled with an equivalent circuit model, containing a Schottky junction diode in parallel with the GIS and MIS capacitors (see inset of Fig. 3a). The capacitance due to the metal-Si contact (bulk contact) is neglected as the contact is believed to be ohmic. For forward bias voltages ($V_{bias}$) above $\Phi_{bi}$ (calculated below), both diodes are in on-state and the depletion width in the Schottky junction is strongly reduced. Therefore, the Schottky junction capacitance is boosted, resulting in a strong increase in the measured capacitance. The decreasing capacitance in the range of -0.6 V to q$\Phi_{bi}$ (compared to Fig. 3a) due to the growing depletion layer capacitance ($C_{depletion}$) of the G/Si junction that overcomes the MIS+GIS capacitance. For $V_{bias}$ < −0.6 V, the bias-independent capacitance shows that the MIS+GIS junction, which is now in deep inversion, dominates the Schottky diode. Therefore, $V_{th}$ for both devices ($D_1$ and $D_2$) is around -0.6 V.

$\Phi_{bi}$, $SBH$ and $N_D$ of the devices can be determined from the measured C-V data. Generally, the extracted $SBH$ is expected to be equal to the $SBH$ estimated from J-V measurements. The estimated value for $N_D$ should be in the same range as the doping density of the Si substrate used to fabricate the devices. The relation between C and $V_{bias}$ can be expressed by Eq. 1 for a non-ideal Schottky diode with an ideality factor $n$ and an interfacial oxide layer between the metal and the semiconductor [34]:

$$\frac{1}{C^2} = \frac{2n[n\left(\Phi_{bi} - \frac{k_BT}{q}\right) - V_{bias}]}{q\varepsilon_s N_D}, \quad (1)$$

where $q$, $T$, and $k_B$ are the elementary charge, the absolute temperature and the Boltzmann constant, respectively. Hence, a linear graph is expected when plotting $1/C^2$ versus $V_{bias}$ for a Schottky junction with a constant $N_D$ within the depletion layer. In addition, the SBH can be obtained using Eq. 2 [34]:

$$SBH = q(\frac{\Phi_{bi}}{n} + \frac{k_B T}{q} + \Phi_n - \Delta\Phi), \quad (2)$$

where $\Delta\Phi$ is the effect of SBH lowering and $\Phi_n$ is the potential difference between the conduction band and Fermi level of Si. $\Phi_n$ can be estimated using Eq. 3 [33]:

$$\Phi_n = k_B T ln\frac{N_C}{N_D}, \quad\quad\quad (3)$$

where $N_C$ is the effective density of the states in the conduction band of Si, which is equal to $2.8 \times 10^{19}$ cm$^{-3}$.

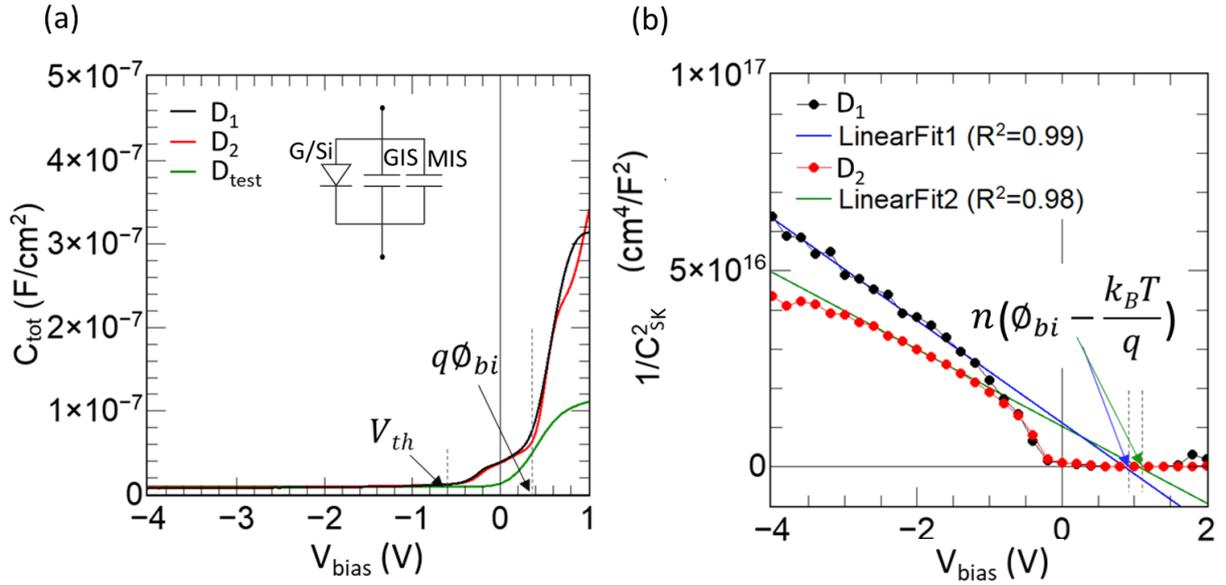

Fig. 3: (a) C-V characteristics of $D_1$ (black) and $D_2$ (red) compared to $D_{test}$ (green). Inset: Equivalent circuit model for the devices ($D_1$ and $D_2$). (b) $1/C^2_{SK}$ vs. $V_{bias}$ of $D_1$ (black) and $D_2$ (red). The intercept of the plot with the $V_{bias}$-axis is equal to $n(\Phi_{bi} - \frac{k_B T}{q})$.

Devices $D_1$ and $D_2$ feature combined G/Si, GIS and MIS structures. Thus, the contribution of the GIS and MIS capacitors must be deducted from the measured total capacitance in order to obtain the capacitance of the Schottky junction (G/Si). Afterwards, $\Phi_{bi}$ and $N_D$ can be extracted from the intercept on the $V_{bias}$-axis and the slope, respectively, of the "1/C$^2$ vs. V$_{bias}$" plot. Then, the SBH can be calculated using Eq. 2. To obtain the C-V characteristics of the Schottky junction, we assumed that the MIS and GIS capacitors of the present devices behave similar to that of the test structure. Therefore, the Schottky junction capacitance for D$_1$ and D$_2$ can be estimated with Eqs. 4 and 5:

$$C_{SK1} = C_{tot1} - \frac{A_1}{A_{D_{test}}} \times C_{D_{test}}, \qquad (4)$$

$$C_{SK2} = C_{tot2} - \frac{A_2}{A_{D_{test}}} \times C_{D_{test}}, \qquad (5)$$

where $C_{SK1}$ ($C_{SK2}$) is the Schottky junction capacitance of D$_1$ (D$_2$), $A_1$ ($A_2$) is the area of GIS+MIS area in device D$_1$ (D$_2$), $A_{D_{test}}$ is the total area of the test device, which is equal to the graphene area plus area of the metal electrode connected to graphene, and $C_{D_{test}}$ is the measured capacitance of the test structure.

Fig. 3b shows the "1/C$^2_{SK}$ vs. V$_{bias}$" plots for both devices D$_1$ and D$_2$. It exhibits the linear characteristics expected for a Schottky junction. $N_D$ of Si as estimated from the slopes of the plots, is $2 \times 10^{15}$ cm$^{-3}$ and $2.7 \times 10^{15}$ cm$^{-3}$ for D$_1$ and D$_2$, respectively. These values are within the range of the doping concentration specified for the Si wafer (between $2 \times 10^{14}$ cm$^{-3}$ and $4 \times 10^{15}$ cm$^{-3}$). Using Eq. 3, we calculated $\Phi_n = 0.24$ V for D$_1$ and $\Phi_n = 0.23$ V for D$_2$. Moreover, $\Phi_{bi}$ values of 0.44 V for D$_1$ and 0.47 V for D$_2$ can be extracted from the intercept on the V$_{bias}$-axis in Fig. 3b. Eq. 2 then yields *SBH* values of 0.68 eV and 0.70 eV for D1 and D2, respectively. Finally, the corresponding ideality factors were determined from the J-V measurements. Here, we have neglected the *SBH* lowering effect ($\Delta\Phi$) in our calculations since the reverse saturation current densities are almost constant in both diodes (see Fig.1c). It should be noted that the *SBH*s determined from the C-V measurements are about 0.1 eV less than those extracted from the J-V measurements. We attribute this small difference to the literature value of the $A^{**}$ (112 Acm$^{-2}$K$^{-2}$) considered when calculating the *SBH*s from the J-V measurements. As pointed out earlier, the $A^{**}$ for G/Si junctions are expected to be smaller than the theoretical value considered and that will result in a smaller *SBH*. We therefore believe that this small difference between the SBHs extracted from I-V and C-V measurements would vanish when using the correct value of $A^{**}$, which is not known at the moment.

**Conclusions**

C-V measurements were performed on G/Si Schottky diodes with inherent parallel GIS and MIS capacitors. We introduced a method for precisely evaluating the C-V characteristics of these structures. The method provides an alternative way of determining the Schottky barrier height and can be used to crosscheck the consistency of values obtained from analyses of measured I-V data, which may be prone to errors because of an unclear exact value of the Richardson constant.

**Methods**

**Device Fabrication:** A phosphorus-doped Si wafer with a resistivity of 1 to 20 $\Omega$.cm was used as the initial substrate. Then, 20 nm $SiO_2$ has been grown on the Si surface using a thermal oxidation process and the wafer has been diced into 20×20 $mm^2$ chips. The n-Si substrate has then been partially exposed by etching the $SiO_2$ layer using buffered oxide etchant (BOE) after a first UV-photolithography step. This step has not been performed for the fabrication of the test structures. Afterwards, metal electrodes have been formed through a second photolithography step, followed by the deposition of (15/85 nm) chromium (Cr)/nickel (Ni) and a liftoff process. Large-area graphene has been grown on a copper (Cu) foil in a Moorfield NanoCVD rapid thermal processing tool using the chemical vapor deposition (CVD) method.[35] A polymer-supported wet chemical etching transfer technique has been used to transfer graphene films onto the pre-patterned substrates immediately after native oxide etching in BOE to ensure a pure G/Si Schottky junction for the diodes.[36] During transfer, the graphene on Cu foil has been spin-coated with Poly methyl methacrylate (PMMA) and baked for 5 minutes at 85°C. After cutting it into ~ 7 $mm^2$ pieces, Iron (III) chloride ($FeCl_3$) has been used to etch the Cu foil and the polymer-supported graphene films have been rinsed in HCL followed by DI-water. Then, the chips have been baked on a hotplate for 35 min at 180°C, followed by removal of the PMMA layer in hot acetone (55°C) for one hour. Afterwards, they have been cleaned with isopropanol and dried with nitrogen. Finally, graphene has been patterned through a last photolithography step, followed by etching of graphene in oxygen plasma.

**Electrical Characterization: I-V measurements** have been performed on the diodes using a Karl Süss probe station connected to a Keithley semiconductor analyzer (SCS4200) under ambient condition. The voltage has been swept from 0 to +2 V for forward ($V_F$) and from 0 to -2V for reverse ($V_R$) biasing, for all the diodes. **C-V measurements** have been carried out using the same probe station connected to an HP Impedance Analyzer (HP 4294A). The measurements have been performed using a small AC signal amplitude of 25 mV at 10 kHz frequency superimposed to a DC bias voltage, which was swept from −4 V to 4 V. The parallel circuit model (conductance in parallel with capacitance) has been used for the measurements.

**Acknowledgements**

Financial support from the European Commission (Graphene Flagship, 785219) and the German Ministry of Education and Research, BMBF (GIMMIK, 03XP0210) is gratefully acknowledged.